\documentclass[conference,10pt]{IEEEtran}
\usepackage{amsmath,amssymb,amsfonts, amsthm}
\usepackage{algorithm}
\usepackage{algorithmic}
\usepackage{graphicx}
\usepackage{textcomp}
\usepackage{xcolor}
\usepackage{bm}
\usepackage{bbm}
\usepackage{cite}
\usepackage{caption}
\usepackage{subcaption}

\newenvironment{Proof}[1]{\medskip\par\noindent{\textbf{Proof:}\,}\,#1}{{\mbox{\,$\blacksquare$}\par}}
\newenvironment{ProofOfTh1}[1]{\medskip\par\noindent{\textbf{Proof of Theorem~1}:\,}\,#1}{{\mbox{\,$\blacksquare$}\par}}

\DeclareMathOperator*{\DP}{DP}
\def\BibTeX{{\rm B\kern-.05em{\sc i\kern-.025em b}\kern-.08em
    T\kern-.1667em\lower.7ex\hbox{E}\kern-.125emX}}

\newtheorem{theorem}{Theorem}
\newtheorem{lemma}{Lemma}

\begin{document}

\title{Index-Based Scheduling for a Resource-Constrained Quantum Switch}

\author{Subhankar Banerjee \qquad Stavros Mitrolaris   \qquad Sennur Ulukus\\
\normalsize Department of Electrical and Computer Engineering\\
\normalsize University of Maryland, College Park, MD 20742\\
\normalsize \emph{sbanerje@umd.edu} \qquad  \emph{stavros@umd.edu}  \qquad \emph{ulukus@umd.edu}}

\maketitle

\begin{abstract}
We consider a quantum switch with a finite number of quantum memory registers that aims to serve multipartite entanglement requests among $N$ users. We propose scheduling policies that aim to optimize the average number of requests served per unit time by efficiently utilizing the switch's available memory. To measure the performance of the scheduling policies, we employ the newly introduced metric of \emph{age of entanglement establishment} (AoEE). We formulate the scheduling problem in a restless multi-armed bandit (RMAB) framework. We show that the scheduling of entanglement requests is \emph{indexable}. Subsequently, we find a closed-form expression of the Whittle index for all possible request-age pairs. By modeling the Whittle index of each request as its reward and its cardinality as its cost, we formulate the memory-constrained scheduling problem as a $0$-$1$ knapsack problem and solve it via dynamic programming. Furthermore, we consider two low-complexity sequential greedy policies that leverage two different modified Whittle indices.
\end{abstract}

\section{Introduction}
The quantum model of computation \cite{nielsen2010quantum} has given rise to a number of applications that rely on establishing entanglement among users across a network. Notable examples of such applications include quantum key distribution \cite{ekert1991quantum}, distributed quantum computation \cite{barral2025review}, entanglement-assisted communication \cite{bennett1999entanglement}, and quantum sensing \cite{zhang2021distributed}. Although one can prepare an entangled quantum state locally and distribute its qubits to remote parties, thus establishing entanglement directly, such an approach cannot be adopted when faced with long-distance transmissions. This is because when transmitting photonic qubits through optical fiber, the probability of successful transmission decays exponentially with distance \cite{azuma2023quantum}. In order to support applications of this kind, quantum switches that leverage entanglement swapping have been proposed. 

\begin{figure}[t]
    \centering
    \begin{subfigure}[t]{0.27\linewidth}
        \centering
        \includegraphics[width=\linewidth]{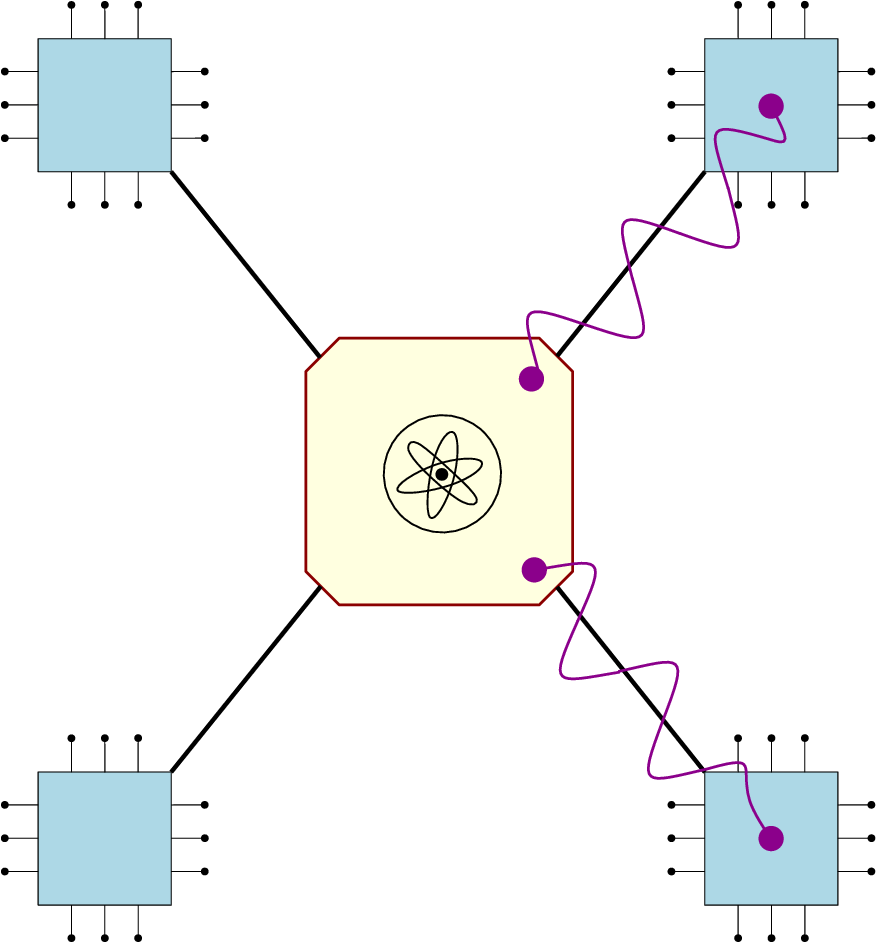}
        \caption{}
    \end{subfigure}\hfill
    \begin{subfigure}[t]{0.27\linewidth}
        \centering
        \includegraphics[width=\linewidth]{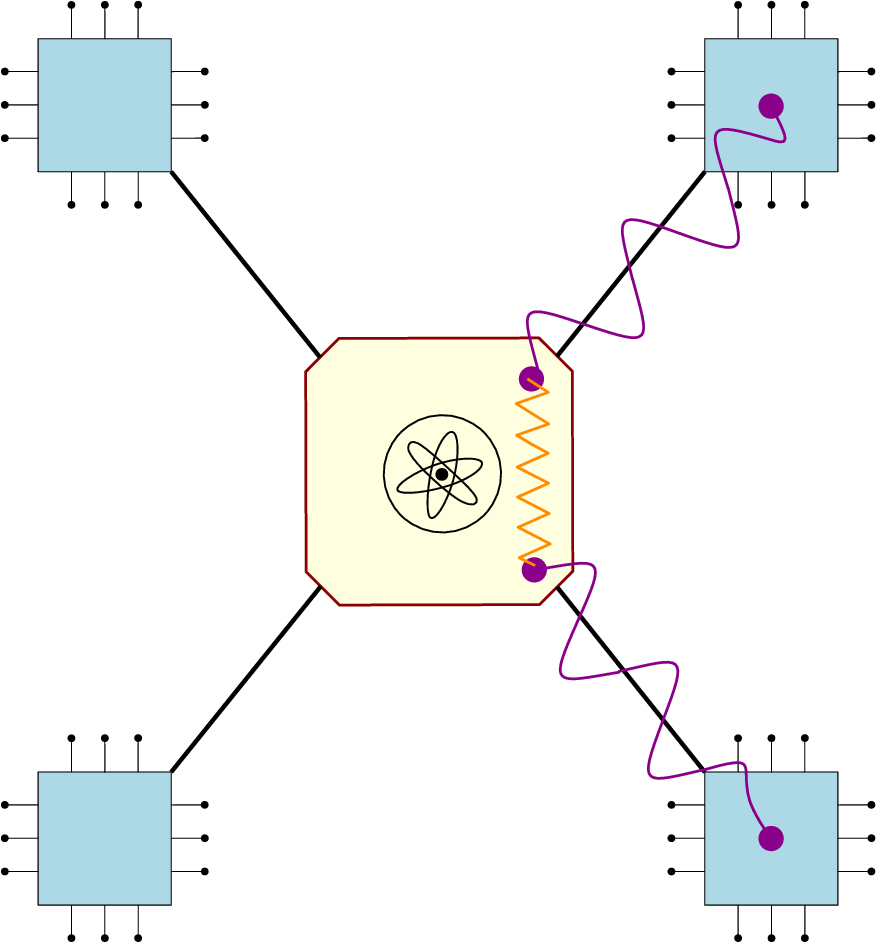}
        \caption{}
    \end{subfigure}\hfill
    \begin{subfigure}[t]{0.27\linewidth}
        \centering
        \includegraphics[width=\linewidth]{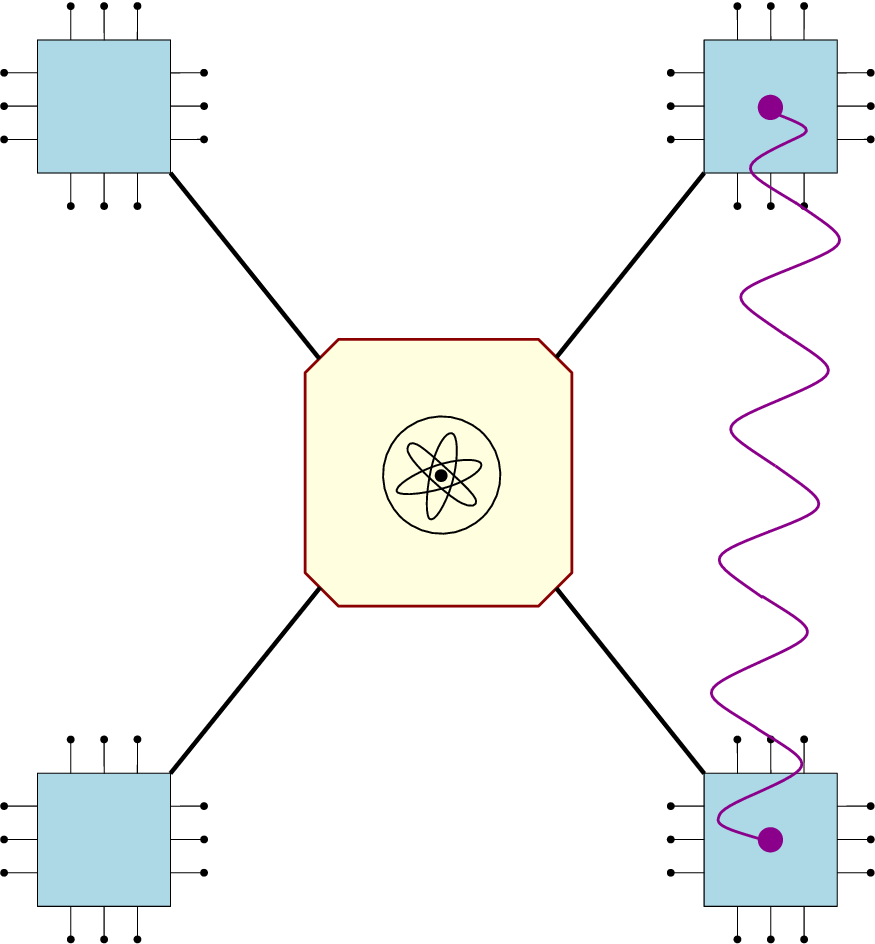}
        \caption{}
    \end{subfigure}
    \caption{An illustrative example of an entanglement swapping operation. In (a), the switch has established LLEs with two users and stores the corresponding qubits in the memory. In (b), the switch performs a Bell-state measurement on the locally stored qubits. In (c), end-to-end entanglement between the users has been established.}
    \label{fig:swapping}
\end{figure}

The primary purpose of a quantum switch is to facilitate the generation of entangled quantum states between remote parties. This is achieved by following a two-step process. First, the switch establishes link-level entanglement (LLE) with users connected to it. An LLE between the switch and a user is a pair of entangled qubits, with one qubit stored at the switch and the other at the user. Assuming that the parties requesting to share entanglement through the switch have all successfully established LLEs with it, the switch performs a local operation on its stored qubits, referred to as entanglement swapping, that, if successful, converts these LLE pairs into end-to-end entanglement among the parties. For two users, the swapping operation corresponds to a Bell-state measurement, whereas for more than two users, it corresponds to a Greenberger-Horne-Zeilinger basis measurement \cite{vardoyan2019stochastic, nielsen2010quantum}. An illustrative example is shown in Fig.~\ref{fig:swapping}. In this work, we consider a memory-constrained quantum switch that aims to serve multipartite entanglement requests. We consider probabilistic LLE generation and swapping operations, as well as one-slot decoherence. 

With quantum switches being important components of quantum networks, researchers have focused both on analyzing their performance and designing control protocols. On the analysis side, \cite{vardoyan2019stochastic, vardoyan2020exact, nain2020analysis} consider a fixed policy for the switch and derive closed-form expressions for the number of end-to-end entanglements served per unit time, and the expected number of quantum memory registers required. On the control side, \cite{dai2022capacity, vasantam2022throughput, promponas2024maximizing, bhambay2025optimal} focus on the problem of queue stability and develop throughput-optimal policies under different network settings. Beyond rate and stability objectives, \cite{panigrahy2023capacity, jia2024fidelity} also incorporate entanglement quality into policy design by accounting for the fidelity of the generated quantum states.

Inspired by the well-known age of information metric \cite{YatesSurvey}, in \cite{mitrolaris2025age}, we proposed a metric coined \emph{age of job completion} to characterize the job timeliness for latency-sensitive jobs. We can think of serving an entanglement request as serving a job. This observation enables us to introduce the notion of age in quantum networks. As a consequence, tools commonly used in the age literature can be repurposed for quantum networking problems. On the other hand, quantum networking introduces new challenges for the community working on timeliness. 

As a first step in this direction, in \cite{mitrolaris2026age}, we introduced the \emph{age of entanglement establishment} metric to evaluate the performance of scheduling algorithms for quantum switches. Since then, quantum networking community has increasingly utilized age-based metrics, e.g., \cite{ercetin2026fidelity, ceran2026age}, that rely on the same fundamental concept of tracking the time elapsed since successful entanglement generation. As stated in \cite{mitrolaris2025age}, minimizing the age of job completion is equivalent to maximizing the average number of completed jobs per unit time. Similarly, it is intuitive to expect that the age of entanglement establishment metric not only introduces a measure of timeliness to the scheduling problem in quantum networks but also inherently increases the number of entanglement requests served per unit time. We note that the direct application of the age of information on quantum networks has been considered in \cite{badia2023strategic, jabrayilova2025age}. However, they consider strategic communication systems and game-theoretic equilibria, rather than designing a specific quality of service metric for scheduling in quantum networks. 

In \cite{mitrolaris2026age}, we proposed three scheduling policies based on randomization and the max-age decision rule. In this work, we formulate the entanglement request scheduling problem as a restless multi-armed bandit (RMAB) problem \cite{verlupewhittle}, where we consider each request as an arm. In \cite{whittleindex}, the author introduced the Whittle index policy to solve problems with RMAB formulations. With certain technical conditions, the Whittle index policies are known to be asymptotically optimal \cite{verlupewhittle}. In related RMAB problems with knapsack-type resource constraints, knapsack-aided index-based policies have been proposed as heuristics; in special cases, they are shown to be optimal \cite{graczova2014generalized, jacko2016resource}. In this work, we first show that our scheduling problem is indexable; subsequently, we derive the explicit Whittle index for each request-age pair. Leveraging these indices as rewards and request cardinalities as costs, we formulate the scheduling problem as a $0$-$1$ knapsack problem. To solve this, we propose a dynamic programming-based approach, which we call the \emph{knapsack-aided Whittle index} (KAWI) policy. Furthermore, we propose two sequential scheduling policies based on modifying the Whittle indices. Finally, we numerically evaluate and compare the performance of our proposed policies.   

\section{System Model}
We consider a time-slotted system with a quantum switch and $N$ users, where the users are arranged in a star topology. We assume that the quantum switch can store at most $M$ LLEs per time slot, constrained by the limited number of available quantum memory registers. We denote the set of end-to-end entanglement establishment requests with $\mathcal{R}=\{1,2,\cdots,R\}$. Each request $i\in\mathcal{R}$ is associated with a specific subset of users, denoted by $\mathcal{G}(i)$, who seek to share entanglement among them. We define the cardinality of request $i$ as the cardinality of the set $|\mathcal{G}(i)|$. We denote the set of all possible distinct cardinalities with $\Lambda$, i.e., $\Lambda=\{|\mathcal{G}(i)|: i\in\mathcal{R}\}$. Note that there can be multiple requests with the same cardinality $\lambda\in\Lambda$; we denote the set of requests with cardinality $\lambda$ with $C(\lambda)$, i.e., $C(\lambda)=\{i\in\mathcal{R}: |\mathcal{G}(i)|=\lambda\}$.

The memory constraint restricts the scheduling of requests to certain subsets of $\mathcal{R}$ whose aggregate cardinality does not exceed $M$ at any given time slot $t$; we refer to such a subset as an admissible set of requests. A scheduling policy $\pi$ selects an admissible set of requests at each time slot $t$, which we denote with $\bar{\mathcal{R}}^{\pi}(t)$. We define the indicator variable $u^{\pi}_{i}(t)$ to denote whether request $i$ is being scheduled at time $t$ by policy $\pi$. Specifically, if $i\in\bar{\mathcal{R}}^{\pi}(t)$, we set $u^{\pi}_{i}(t)=1$, and we set $u^{\pi}_{i}(t)=0$ otherwise. If $u^{\pi}_{i}(t)=1$,  $|\mathcal{G}(i)|$ memory registers are allocated for request $i$, allowing all  users in $\mathcal{G}(i)$ to attempt to establish an LLE  with the switch. We assume that user $j$ successfully establishes an LLE with the switch with probability $p_{j} \in(0,1]$, independently across time and of other users' LLE establishment. Note that a user $j$ may participate in multiple requests; thus, it is possible that at time $t$, user $j$ establishes multiple LLEs with the switch, one for each request. We define the indicator variable $b_{i,j}^{\pi}(t)=1$ if user $j$ successfully establishes an LLE for request $i$ at time $t$, and $0$ otherwise. Thus, given that request $i$ is scheduled at time $t$, i.e., $u_{i}^{\pi}(t)=1$, the probability that $b_{i,j}^{\pi}(t)$ takes the value $1$ is $p_{j}$, for all $j\in\mathcal{G}(i)$.

\begin{figure}[t]
    \centerline{\includegraphics[width = 0.9\columnwidth]{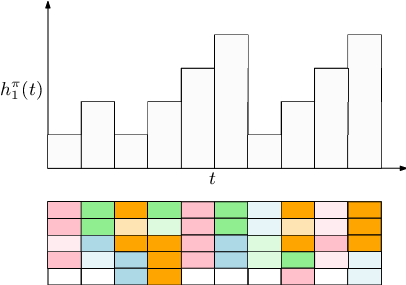}}
    \caption{Example with $R=4$ requests, with $|\mathcal{G}(1)|=2, |\mathcal{G}(2)|=2$, $|\mathcal{G}(3)|=3$, and $|\mathcal{G}(4)|=4$. The requests are represented by the colors green, blue, orange, and pink, respectively. The top figure shows the evolution of the age of entanglement establishment for request $1$ (green) under policy $\pi$, and the bottom figure shows the corresponding memory-register allocation with $M=5$ memory registers. At time $t$, a memory block allocated to a request is shown using a lighter shade of the corresponding color, while a successfully established LLE for that request is shown using the darker shade. For example, at time slot $1$, request $4$ (pink) is scheduled, and the first four blocks are allocated to it; however, only three users in the request successfully establish LLEs.}
    \label{fig:1}
\end{figure}

For request $i$, if all the users in the set $\mathcal{G}(i)$ successfully establish LLEs with the switch, we define the indicator variable $c_{i}^{\pi}(t)$ to be $1$, and $0$ otherwise. Note that, $c_{i}^{\pi}(t)= \prod_{j\in\mathcal{G}(i)} b_{i,j}^{\pi}(t)$; thus, given that $u_{i}^{\pi}(t)=1$, we have $c_{i}^{\pi}(t)=1$ with probability $v(i) = \prod_{j\in\mathcal{G}(i)} p_{j}$. If $u_{i}^{\pi}(t)=1$ and $c_{i}^{\pi}(t)=0$, it implies that at least one user in $\mathcal{G}(i)$ failed to establish an LLE. However, some users associated with request $i$ may still have successfully established LLEs. We assume that those LLEs are unused and dropped at the end of time $t$. A system where such LLEs are repurposed to serve other requests would effectively improve performance; therefore, our results serve as an upper bound on the achievable age for such systems. For request $i$, if $c_{i}^{\pi}(t)=1$, the switch attempts to generate end-to-end entanglement among all users in the set $\mathcal{G}(i)$. This operation succeeds with probability $q_{\lambda} \in (0,1]$, where $\lambda = |\mathcal{G}(i)|$. We denote the outcome by the indicator variable $d_{i}^{\pi}(t)$, which takes the value $1$ if successful and $0$ otherwise. Thus, we have 
\begin{align}\label{eq:1}
    d^{ \pi}_i(t) = 
    \begin{cases}
        0,  & \text{if } u^{ \pi}_i(t) =0,  \\
        0, & \text{if } u^{ \pi}_i(t) =1 \: \text{ w.p.} \; 1-q_\lambda v(i) ,\\
        1, & \text{if } u^{ \pi}_i(t) =1 \: \text{ w.p.} \;  q_\lambda v(i).
    \end{cases}
\end{align}
Following \cite{mitrolaris2025age}, we define the age of entanglement establishment $h_{i}^{\pi}(t)$ for request $i$ at time $t$ as the time elapsed since the last successful establishment of end-to-end entanglement among the users in $\mathcal{G}(i)$. Formally,
\begin{align}
    h^{ \pi}_i(t) = t - \sup\{ t' \in \{1,2,\ldots,t-1\}:  \; d^{ \pi}_i(t') = 1 \},
\end{align}
where the supremum of an empty set is defined to be $0$. In Fig.~\ref{fig:1}, we pictorially represent the evolution of the age of entanglement establishment.

In this work, we aim to design policies that minimize the long-term average age of entanglement subject to a limited number of quantum memory registers $M$. Let $\Pi$ denote the set of all causal scheduling policies.  Specifically, we consider the following optimization problem,
\begin{align}
    &\inf_{\pi\in\Pi} \limsup_{T\rightarrow\infty} \frac{1}{T}\frac{1}{R}\sum_{i=1}^{R}\sum_{t=1}^{T} \mathbb{E}[h_{i}^{\pi}(t)] \\& \ \textrm{s.t.}\   \sum_{i=1}^{R} |\mathcal{G}(i)| u_{i}^{\pi}(t) \leq {M}, \quad \forall t\in\{1,2,\cdots\}. \label{eq:n2}
\end{align}

\section{KAWI Policy}\label{sec:kanpsack}
In this section, we propose a Whittle index policy that uses dynamic programming to schedule the requests. First, we relax the per-slot constraint in (\ref{eq:n2}) to an expected time average constraint, thus the relaxed problem becomes,
\begin{align}
    &\inf_{\pi\in\Pi} \limsup_{T\rightarrow\infty} \frac{1}{T}\frac{1}{R}\sum_{i=1}^{R}\sum_{t=1}^{T} \mathbb{E}[h_{i}^{\pi}(t)] \label{eq:15}\\& \ \textrm{s.t.} \limsup_{T\rightarrow\infty}\frac{1}{T} \frac{1}{R}\sum_{i=1}^{R}\sum_{t=1}^{T} |\mathcal{G}(i)| \mathbb{E}[u_{i}^{\pi}(t)] \leq \frac{M}{R}. \label{eq:16}
\end{align}
We employ a Lagrange multiplier $\tau\geq 0$, to form the following Lagrangian corresponding to the problem in (\ref{eq:15})-(\ref{eq:16}),
\begin{align}
    L(\pi;\tau) = &\limsup_{T\rightarrow\infty} \frac{1}{RT}\Big(\sum_{i=1}^{R}\sum_{t=1}^{T} \mathbb{E}[h_{i}^{\pi}(t) + \tau |\mathcal{G}(i)| u_{i}^{\pi}(t)]\Big) \nonumber\\&-\tau \frac{M}{R}.
\end{align}
Now, consider the following problem,
\begin{align}
    \inf_{\pi\in\Pi} L(\pi;\tau),\label{eq:18}
\end{align}
First, note that the constant $\tau\frac{M}{R}$ is independent of $\pi$, thus we omit it for the problem in (\ref{eq:18}). Also, note that the problem (\ref{eq:18}) is linearly separable across different requests. Thus, for a given $\tau\geq 0$, (\ref{eq:18}) can be partitioned into $R$ different sub-problems. The sub-problem for the $i$th request becomes,
\begin{align}
    \inf_{\pi\in\Pi} \limsup_{T\rightarrow\infty} \frac{1}{T} \sum_{t=1}^{T} \mathbb{E}[h_{i}^{\pi}(t) + \tau |\mathcal{G}(i)|u_{i}^{\pi}(t)].\label{eq:19}
\end{align}
Recall that $u_{i}^{\pi}(t)$ is an indicator random variable. In (\ref{eq:19}), at time $t$, if $u_{i}^{\pi}(t)=1$, we pay a scheduling cost $\tau |\mathcal{G}(i)|$, while choosing $u_{i}^{\pi}(t)=0$ does not require us to pay any such cost. Following the literature, we call $u_{i}^{\pi}(t)=1$ as the \emph{active action} and $u_{i}^{\pi}(t)=0$ as the \emph{passive action}. Note that, we can think of (\ref{eq:19}) as a Markov decision process (MDP) with space $\mathcal{S}_{i}$, where a state $s\in\mathcal{S}_{i}$ is the age of request $i$. Thus, at time $t$ for a policy $\pi$ the age of the $i$th request $h_{i}^{\pi}(t)$ takes a value from the set $\mathcal{S}_{i}$, i.e., there exists a $s\in\mathcal{S}_{i}$, such that $h_{i}^{\pi}(t)=s$.
 
For $\tau\geq 0$, we denote $\mathcal{P}_{i}(\tau)$ as a set consisting of all the states in $\mathcal{S}_{i}$, for which the passive action is optimal. We say that the sub-problem (\ref{eq:19}) corresponding to the $i$th request  is indexable, if for $\tau_{2}\leq \tau_{1}$, we have $\mathcal{P}_{i}(\tau_{2}) \subseteq \mathcal{P}_{i}(\tau_{1})$ and $\mathcal{P}_{i}(\infty)=\mathcal{S}_{i}$. We say that the problem in (\ref{eq:15})-(\ref{eq:16}) is indexable if all sub-problems corresponding to requests $i\in\mathcal{R}$ are indexable. 

\begin{theorem}\label{th:1}
    The scheduling problem in (\ref{eq:15})-(\ref{eq:16}) is indexable.
\end{theorem}
The proof of Theorem~\ref{th:1} is provided in Appendix~\ref{appen:th1}.  
 
Now, we define the Whittle index for a state $s_{i}\in\mathcal{S}_{i}$ as
\begin{align}
    W_{i}(s_{i}) = \inf\{{\tau|s_{i}\in\mathcal{P}_{i}(\tau)}\}.
\end{align}
In the next theorem, we provide the explicit Whittle index expression for state $s_{i}\in\mathcal{S}_{i}$.

\begin{theorem}\label{th:4}
    For request $i \in \mathcal{R}$ with cardinality $\lambda$ and state $s_{i}\in\mathcal{S}_{i}$, we have,
    \begin{align}\label{eq:13}
        W_{i}(s_{i}) = \frac{s_{i}\left(q_{\lambda}v(i)s_{i} - q_{\lambda}v(i) +2\right)}{2 \lambda}.
    \end{align}
\end{theorem}
The proof of Theorem~\ref{th:4} is provided in Appendix~\ref{appen:th2}.

Given the closed-form expression of the Whittle index in Theorem~\ref{th:4}, we now devise a scheduling policy $\bar{{\pi}}$ that satisfies the per-slot memory constraint in (\ref{eq:n2}). At the beginning of time slot $t$, the switch observes the age vector $(h_{i}^{\bar{{\pi}}}(t))_{i=1}^{R}$. For each request $i\in\mathcal{R}$, it evaluates the current Whittle index $W_{i}(h_{i}^{\bar{{\pi}}}(t))$. For notational convenience, we define,
\begin{align}\label{eq:25}
    w_{i}(t) \triangleq |\mathcal{G}(i)|W_{i}(h_{i}^{\bar{{\pi}}}(t)).
\end{align}
At time $t$, we interpret $w_{i}(t)$ as the instantaneous reward of scheduling request $i$, with the memory requirement $|\mathcal{G}(i)|$. Given the memory constraint $M$, at time $t$, we obtain the policy $\bar{{\pi}}$ by solving the $0$-$1$ knapsack problem with dynamic programming. We define the following recursion on $i$,
\begin{align}\label{eq:23}
    \textrm{DP}_{t}(i,c) = 
    \begin{cases}
        \max \big\{ \textrm{DP}_{t}(i-1,c), w_{i}(t) 
        \\ \quad \quad \ + \textrm{DP}_{t}(i-1,c-|\mathcal{G}(i)|) \big\}, &\text{if} \ |\mathcal{G}(i)|\leq c, \\ \textrm{DP}_{t}(i-1,c), & \text{if} \ |\mathcal{G}(i)|>c,
    \end{cases}
\end{align}
where $i$ and $c$ take values from sets $\mathcal{R}$ and $\{0,1,\cdots,M\}$, respectively. We use the following initializations, 
\begin{align}
    \textrm{DP}_{t}(0,c)=&0, \quad \forall{c\in\{0,1,\cdots,M\}}, \\ \textrm{DP}_{t}(i,0) = &0, \quad \forall{i\in\mathcal{R}}.
\end{align}
Once we have $\textrm{DP}_{t}(R,M)$ by evaluating the iteration in (\ref{eq:23}), we obtain the set of requests $\bar{\mathcal{R}}_{t}^{\bar{\pi}}\subseteq\mathcal{R}$ by a backward iterative process described in Algorithm~\ref{alg:1}.  At time $t$, the switch schedules all the requests in the set $\bar{\mathcal{R}}_{t}^{\bar{\pi}}$.

\begin{algorithm}[t]
\caption{Backtracking to obtain Set $\bar{\mathcal{R}}_t^{\bar{\pi}}$}
\label{alg:1}
\begin{algorithmic}[1]
    \STATE $\bar{\mathcal{R}}_t^{\bar{\pi}} \gets \emptyset$
    \STATE $c \gets M$
    \FOR{$i = R$ \textbf{downto} $1$}    
        \IF{$|\mathcal{G}(i)| \leq c$ \AND $\DP_t(i,c) = w_{i}(t) + \DP_{t}(i-1, c - |\mathcal{G}(i)|)$}
            \STATE $\bar{\mathcal{R}}_t^{\bar{\pi}} \gets \bar{\mathcal{R}}_t^{\bar{\pi}} \cup \{ i \}$
            \STATE $c \gets c -|\mathcal{G}(i)|$
        \ELSE
            \STATE  request $i$ is not included; do nothing
        \ENDIF   
    \ENDFOR
    \RETURN $\bar{\mathcal{R}}_t^{\bar{\pi}}$
\end{algorithmic}
\end{algorithm}

\section{Sequential Whittle Index Policy}
In Section~\ref{sec:kanpsack}, we solve the dynamic programming in (\ref{eq:23}) to get the set of scheduled requests $\bar{\mathcal{R}}_{t}^{\bar{\pi}}$ at time $t$, the complexity of which increases linearly with $R$ and $M$. Specifically, the complexity of the dynamic programming in (\ref{eq:23}) is $O(RM)$. Thus, for large $M$, the dynamic programming in (\ref{eq:23}) can be computationally challenging. In this section, we introduce two sequential policies that leverage the Whittle indices derived in (\ref{eq:13}). In contrast to the dynamic programming-based approach, their complexity is independent of the memory size $M$; specifically, they have per-slot complexity $O(R\log{R})$.  

An immediate, low-complexity policy is the greedy policy, which schedules requests sequentially, prioritizing those with higher Whittle indices until the remaining memory can no longer accommodate additional requests. Note that, for the $i$th request, the index in (\ref{eq:25}) does not explicitly depend on the request cardinality $|\mathcal{G}(i)|$. Thus, the cardinalities of two requests can be different from each other; however, if their ages are comparable, then their indices are also comparable. Thus, for two requests $i$ and $i'$ with $|\mathcal{G}(i)|>|\mathcal{G}(i')|$, the same age and the same probability of success, i.e., $q_{|\mathcal{G}(i)|} v(i) = q_{|\mathcal{G}(i')|} v(i')$, the greedy policy gives the same preference to both requests. Consequently, at time $t$, the set of requests scheduled by the greedy policy may differ significantly from the set $\bar{\mathcal{R}}^{\bar{\pi}}(t)$. For example, consider $M=6$, $R=4$ with three bipartite requests and one four-partite request, where all requests have the same probability of success. At time $t$, assume that all four requests have similar ages, with the three bipartite requests having $w_{i}(t)=2.5$ and the four-partite request having $w_{i'}(t)=3$. Thus, at time $t$, the greedy policy schedules the four-partite and one bipartite request. However, the policy described in Section~\ref{sec:kanpsack} schedules the three bipartite requests. Intuitively, if two requests have similar ages, we should prioritize scheduling the request with the lower cardinality, as it requires fewer memory registers. To implement this, in a sequential manner, rather than employing dynamic programming as in Section~\ref{sec:kanpsack}, we consider the following two indices for request $i$,
\begin{align}
    \hat{w}_{i}(t) =& w_{i}(t) - \gamma |\mathcal{G}(i)| , \label{eq:nn16}\\ \tilde{w}_{i}(t) =& \frac{w_{i}(t)}{\beta +  |\mathcal{G}(i)|},  \label{eq:nn17}
\end{align}
where $\gamma$ and $\beta$ are strictly positive constants.
We now consider two policies $\hat{\pi}$ and $\tilde{\pi}$, which schedule requests sequentially by prioritizing them based on the indices $\hat{w}_{i}(t)$ and $\tilde{w}_{i}(t)$, respectively. This sequential scheduling continues until the remaining memory can no longer accommodate additional requests. We call the policy described in (\ref{eq:nn16}) the sequential Whittle index subtractive (SWIS) policy, and the policy described in (\ref{eq:nn17}) the sequential Whittle index divisive (SWID) policy. In Section~\ref{sec:num}, for different network settings, we numerically find the optimal values for $\gamma$ and $\beta$ through a grid search over a given range of values. We then evaluate the performances of the proposed policies numerically with these tuned parameters.  

\section{Numerical Analysis}\label{sec:num}
In this section, we evaluate the performance of our three proposed algorithms under two different network settings. We also compare them with our previously proposed SMW and MMA policies in \cite{mitrolaris2026age}. We use the same simulation settings as in \cite{mitrolaris2026age}. For completeness of this work, we summarize the simulation settings here: We consider a network with $N=5$ users, and assume that the request set $\mathcal{R}$ contains all possible entanglement requests. Thus, the total number of requests is $2^{N}-(N+1) = 26$. We consider the LLE establishment probabilities to be $p_{1}=0.85$, $p_{2}=0.9$, $p_{3}=0.93$, $p_{4}=0.87$, and $p_{5}=0.95$, and the swapping success probabilities to be $q_{2}= 0.92$, $q_{3}=0.87$, $q_{4}= 0.83$, and $q_{5}=0.8$. 

\begin{figure}[t]
    \centerline{\includegraphics[width = 0.925\columnwidth]{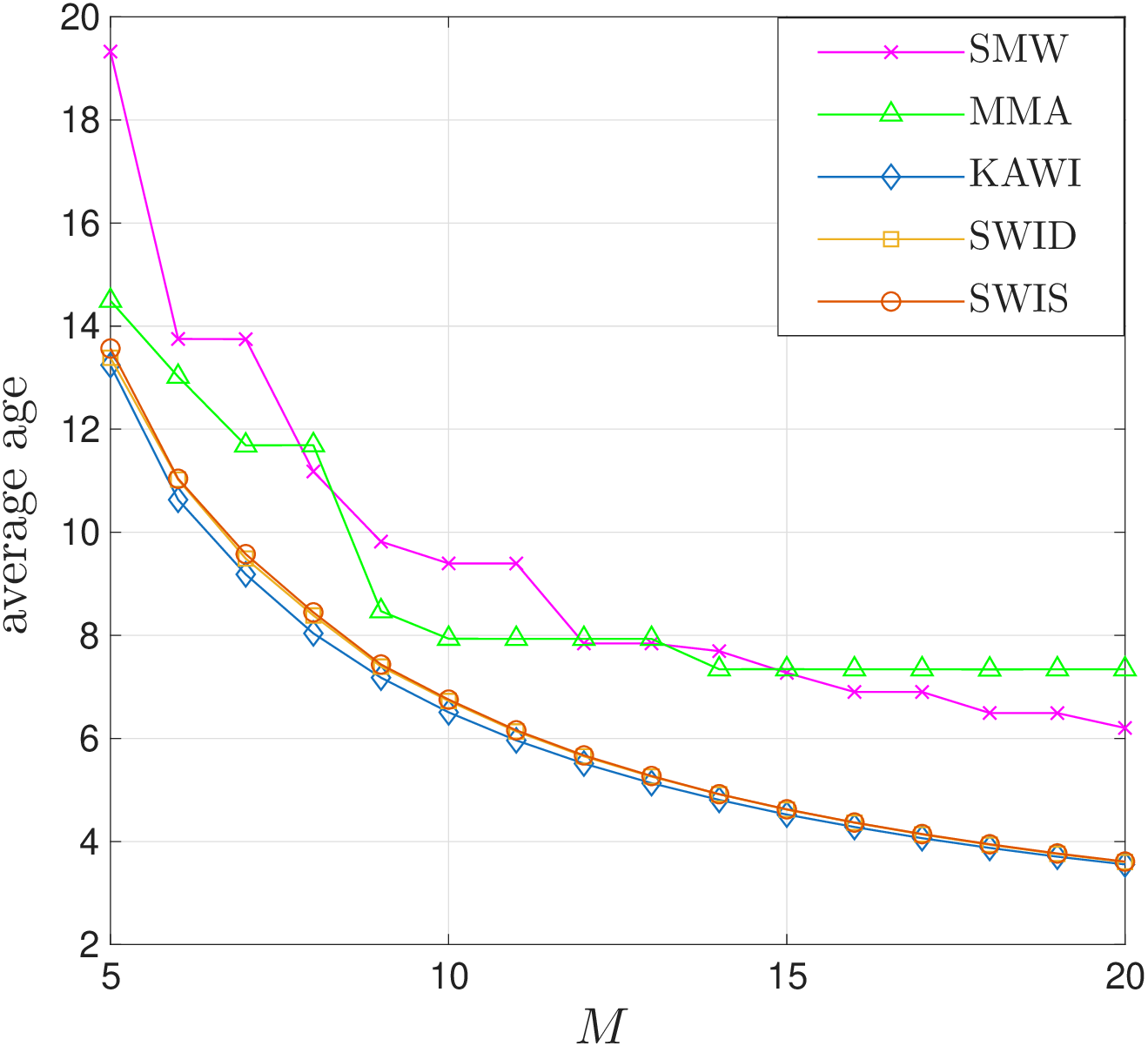}}
    \caption{Average age of entanglement establishment achieved by the proposed policies of this work and of \cite{mitrolaris2026age} as a function of the memory size $M$ in a network with $N=5$ users and all possible requests, $R=2^N-(N+1)=26$.}   
    \label{fig:3}
\end{figure}

In Fig.~\ref{fig:3}, we first study the effect of the varying memory size $M\in\{5,6,\ldots,20\}$ on the age of entanglement establishment. For each value of $M$, we optimize $\gamma$ over the range $0$ to $10$ in increments of $\frac{1}{2}$ for the SWIS policy. Similarly, for each $M$, we optimize $\beta$ over the range $1$ to $10$ in increments of $\frac{1}{2}$ for the SWID policy. From Fig.~\ref{fig:3}, we see that the KAWI policy outperforms all the other policies. However, the SWID and SWIS policies perform very close to the KAWI policies. We also observe that all three proposed policies in this work outperform the policies introduced in \cite{mitrolaris2026age}. Furthermore, we observe that as $M$ increases, the gap between SMW and the three proposed policies remains relatively unchanged with varying $M$. However, the performance gap between MMA and the three proposed policies increases with $M$. This is expected because it is noted in \cite{mitrolaris2026age} that the MMA policy saturates once $M\geq 14$ and no longer improves with increasing $M$. This makes this policy not suitable for systems with a large number of quantum memory registers.

\begin{figure}[t]
    \centerline{\includegraphics[width = \columnwidth]{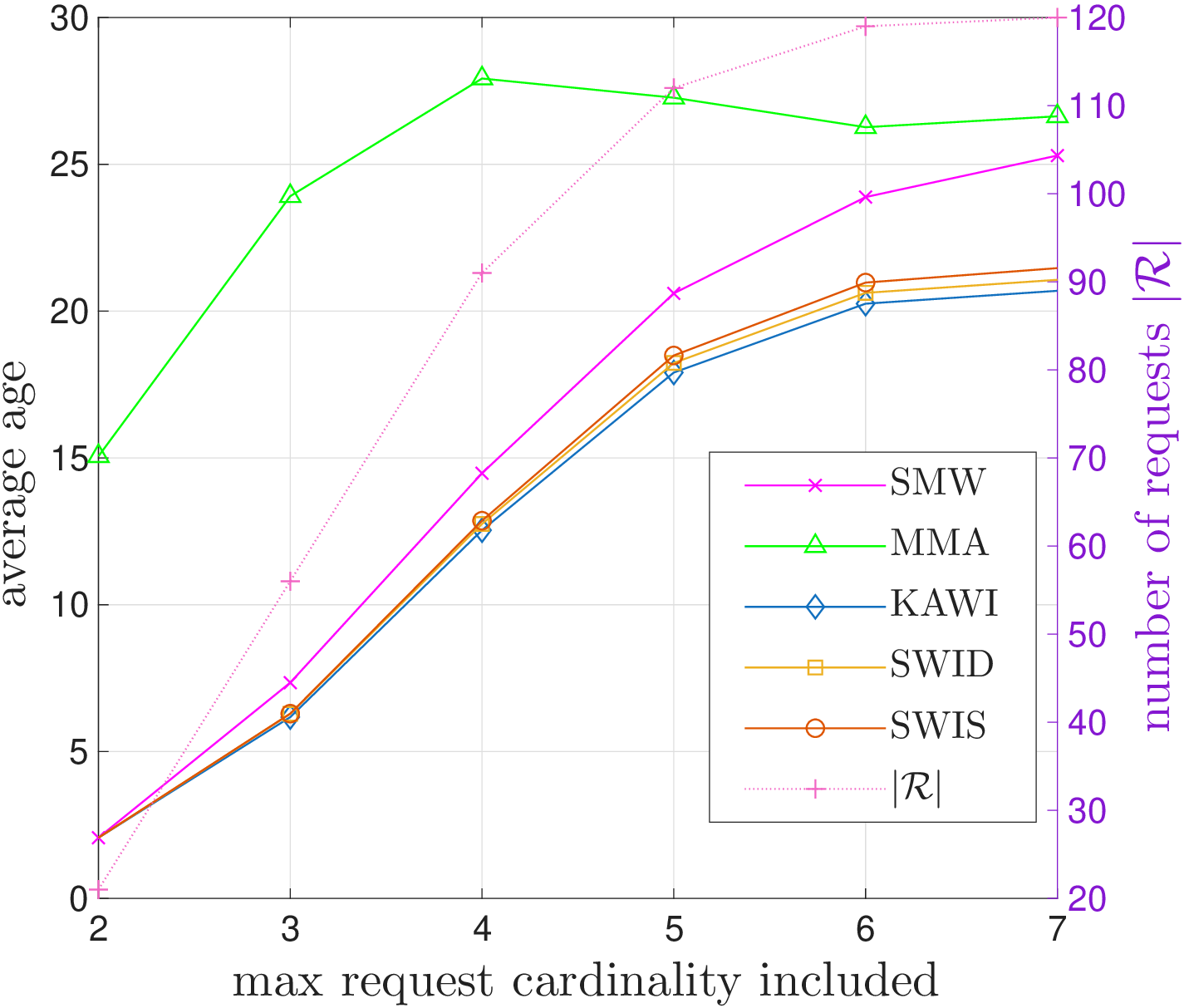}}
    \caption{Average age of entanglement establishment achieved by the proposed policies of this work and of \cite{mitrolaris2026age} as $\mathcal{R}$ expands in a network with $N=7$ users and  $M=20$.}
    \label{fig:4}
\end{figure}

Next, we compare the performances of the policies with varying $\mathcal{R}$ and fixed memory size. We consider a network of $N=7$ users and $M=20$. We consider the LLE establishment probabilities to be $p_1=0.85$, $p_2=0.9$, $p_3=0.93$, $p_4=0.87$, $p_5=0.95$, $p_6=0.83$, and $p_7=0.92$, and the swapping success probabilities to be $q_2=0.92$, $q_3=0.87$, $q_4=0.83$, $q_5=0.8$, $q_6=0.78$ and $q_7=0.75$. In Fig.~\ref{fig:4}, the $x$-axis denotes the maximum cardinality of any request in $\mathcal{R}$. For each value on the $x$-axis, we let $\mathcal{R}$ contain all possible requests whose cardinalities are less than or equal to that value. For example, when $x=4$, the set $\mathcal{R}$ contains all bipartite, all tripartite, and all four-partite requests. Therefore,
\begin{align}
     R = {7 \choose 2} + {7 \choose 3} + {7 \choose 4}.
\end{align}
For each fixed set of requests $\mathcal{R}$, we optimize $\gamma$ and $\beta$ over the same range of values considered earlier. From Fig.~\ref{fig:4}, we observe that KAWI performs the best among all the proposed policies, while all three proposed policies, namely KAWI, SWID, and SWIS, perform close to each other. We also observe that the performance gap between SMW and the three proposed policies increases as we increase the number of requests in $\mathcal{R}$. 

\section{Conclusion}
In this work, we studied the problem of scheduling entanglement requests in a memory-constrained quantum switch, where we measure the performance of a scheduling policy with the age of entanglement establishment (AoEE). We formulated the problem as an RMAB problem, proved the indexability, and derived the closed-form expression for the Whittle index for each request-age pair. Leveraging the Whittle index, we proposed the KAWI policy, which solves a per-slot 0-1 knapsack problem via dynamic programming. Furthermore, we proposed two low-complexity sequential policies, namely, SWIS and SWID, based on modified Whittle indices. Numerical results showed that KAWI achieves the best performance, while SWIS and SWID perform close to KAWI. We also showed that all three proposed policies outperform the previously proposed SMW and MMA policies in the considered network settings. These results show that index-based scheduling policies provide an effective and scalable approach for scheduling requests in a memory-constrained quantum switch.

\appendices

\section{Proof of Theorem~\ref{th:1}}\label{appen:th1}
To prove this theorem, we first mention essential results from the MDP literature \cite{bertsekas2012dynamic, sennott1989average}, to keep this work self-contained. For action $a\in\{0,1\}$, we denote the transition probability from state $s\in\mathcal{S}_{i}$ to $s'\in\mathcal{S}_{i}$ with $P_{a}(s,s')$. Following the dynamics described in (\ref{eq:1}), we list all possible non-zero transition probabilities next.
\begin{align}
    P_{0}(s,s+1) = &1, \nonumber\\
    P_{1}(s,1) = &q_{\lambda} v(i), \nonumber\\
    P_{1}(s,s+1) = &1-  q_{\lambda} v(i). 
\end{align}
where $\lambda$ is the cardinality for the request $i$, i.e., $\lambda=|\mathcal{G}(i)|$. For a state-action pair $(s,a)$ and a given Lagrange multiplier $\tau\geq0$, we denote the immediate cost as $C(s,a;\tau)$. Thus,
\begin{align}
    C(s,a;\tau)=\begin{cases}
        s + \tau|\mathcal{G}(i)|, & \text{if $a=1$}, \\ s, & \text{if $a=0$}.
    \end{cases}
\end{align}
For $0<\alpha<1$, state $s\in\mathcal{S}_{i}$, and a policy $\pi$, we denote the discounted cost as,
\begin{align}
    V_{\alpha}^{\pi}(s;\tau) = \sum_{t=1}^{\infty} \alpha^{t} \mathbb{E}\big[h_{i}^{\pi}(t) + \tau |\mathcal{G}(i)| u_{i}^{\pi}(t) \big].
\end{align}
We denote the discounted value function by
\begin{align}
V_{\alpha}(s;\tau) = \inf_{\pi} V_{\alpha}^{\pi}(s;\tau).
\end{align}
From the literature, we know that
\begin{align}
    V_{\alpha}(s;\tau) = &\min_{a\in\{0,1\}} V_{\alpha}(s,a;\tau),\label{eq:21} \\ V_{\alpha}(s,a;\tau) = &C(s,a;\tau) + \alpha\sum_{s'\in\mathcal{S}_{i}} P_{a}(s,s') V_{\alpha}(s';\tau).
\end{align}
From \cite{bertsekas2012dynamic}, we say that the action that achieves the minimization in (\ref{eq:21}) is optimal, for state $s$, corresponding to the discounted cost criterion. Now, consider the following iteration for $n\geq 1$, 
\begin{align}
    V_{\alpha,n}(s;\tau) = &\min_{a\in\{0,1\}} V_{\alpha,n}(s,a;\tau), \label{eq:24}\\ V_{\alpha,n}(s,a;\tau) = &C(s,a;\tau) +\alpha \! \sum_{s'\in\mathcal{S}_{i}} \!\!P_{a}(s,s') V_{\alpha,n\!-\!1}(s';\tau),\!\!
\end{align}
with $V_{\alpha,0}(s;\tau)=0$, for all $s\in\mathcal{S}_{i}$. From \cite{sennott1989average}, we know that,
\begin{align}\label{eq:n25}
    V_{\alpha}(s;\tau) = \lim_{n\rightarrow\infty} V_{\alpha,n}(s;\tau).
\end{align}

First, we state and prove the next lemma that studies a monotonicity property of the value function in (\ref{eq:21}) with respect to $s$.
\begin{lemma}\label{lemma:1}
    For a given $\tau$, $V_{\alpha}(s;\tau)$ is an increasing function of $s$.
\end{lemma}

\begin{Proof}
    We first show that $V_{\alpha,n}(s;\tau)$ is an increasing function of $s$, for all $n\in \mathbb{N}$, by mathematical induction. Then, from (\ref{eq:n25}), this lemma follows. 
    
    For $n=0$, $V_{\alpha,0}(s;\tau)=0$ for all $s\in\mathcal{S}$, which makes the statement of this lemma immediate for $n=0$. We assume that $V_{\alpha,n}(s;\tau)$ is an increasing function of $s$. Now,
    \begin{align}
        V_{\alpha,n+1}(s,0;\tau) = &s + \alpha V_{\alpha,n}(s+1;\tau),\label{eq:27} \\ V_{\alpha,n+1}(s,1;\tau) = &s+  \tau |\mathcal{G}(i)|+\alpha \Big(q_{\lambda} v(i) V_{\alpha,n}(1;\tau) \nonumber\\&+(1-q_{\lambda}v(i) )V_{\alpha,n}(s+1;\tau)\Big).\label{eq:28}
    \end{align}
    From (\ref{eq:27}), (\ref{eq:28}), and the $n$th induction step we observe that $V_{\alpha,n+1}(s,0;\tau)$ and  $V_{\alpha,n+1}(s,1;\tau)$ are increasing functions of $s$. Thus, from (\ref{eq:24}) we conclude that $V_{\alpha,n+1}(s;\tau)$ is an increasing function of $s$. 
\end{Proof}

In the next lemma, we establish a threshold structure of an optimal policy for the problem in (\ref{eq:19}).

\begin{lemma}\label{lemma:2}
 Consider the $i$th sub-problem in (\ref{eq:19}). For a given $\tau$, if scheduling request $i$, i.e., if action $a=1$ is optimal for state $s$, then action $a=1$ is also optimal for state $s+x$, for $x\geq0$.
\end{lemma}

\begin{Proof}
    First, we show that the threshold structure holds for the discounted cost criterion. Then, following the methodology of \cite{sennott1989average}, we can show that the same structure remains optimal for the average cost formulation in (\ref{eq:19}). 
    
    To prove this lemma for the discounted cost criterion, we show that the following holds for $0<\alpha<1$ and $s\in\mathcal{S}_{i}$,
    \begin{align}\label{eq:29}
        V_{\alpha}(s,1;\tau) -V_{\alpha}(s,0;\tau) \geq V_{\alpha}(s+x,1;\tau) - V_{\alpha}(s+x,0;\tau).
    \end{align}
    Rearranging (\ref{eq:29}), we have
    \begin{align}\label{eq:30}
        &V_{\alpha}(s+x+1;\tau) - V_{\alpha}(s+1;\tau) \nonumber\\& \qquad \!\geq\! (1\!-\!q_{\lambda}v(i)) \big(V_{\alpha}(s+x+1;\tau) - V_{\alpha}(s+1;\tau)\big).
    \end{align}
    From Lemma~\ref{lemma:1}, for $x\geq 0$, we have, 
    \begin{align}
        V_{\alpha}(s+x+1;\tau) - V_{\alpha}(s+1;\tau) \geq 0.
    \end{align}
    Thus, (\ref{eq:30}) holds, which concludes the proof of Lemma~\ref{lemma:2}.
\end{Proof}

Now, we proceed with the proof of Theorem~\ref{th:1}.

\begin{ProofOfTh1}
    We show that an arbitrary request $i\in\mathcal{R}$ is indexable, and the indexability of (\ref{eq:15}) and (\ref{eq:16}) follows directly from the definition of indexability. Thus, we show that for two distinct Lagrangian multipliers $\tau_{1}$ and $\tau_{2}$, where $\tau_{2}\leq \tau_{1}$, we have $\mathcal{P}_{i}(\tau_{2}) \subseteq \mathcal{P}_{i}(\tau_{1})$ and $\mathcal{P}_{i}(\infty)=\mathcal{S}_{i}$. For a fixed Lagrangian multiplier $\tau$, we first find the average cost of an optimal policy for the MDP in (\ref{eq:19}).  From Lemma~\ref{lemma:2}, we know that there exists an optimal policy for (\ref{eq:19}) that has a threshold structure. We denote a threshold with $\Delta\geq 1$, and we denote the corresponding policy as $\pi_{\Delta}$. Under the policy $\pi_{\Delta}$, the $i$th request is not scheduled until its age reaches the threshold $\Delta$. Thereafter, the $i$th request is scheduled in every slot until the end-to-end entanglements corresponding to the $i$th request are successfully established and the age of entanglement establishment for the $i$th request drops to $1$. In other words, for the policy $\pi_{\Delta}$, action $a=0$ is taken for states $s<\Delta$, whereas action $a=1$ is taken for states $s\geq\Delta$. 

    For the policy $\pi_{\Delta}$, we divide the whole time horizon $T$ into consecutive frames, where a frame ends when the $i$th request gets served successfully, and the next frame begins immediately thereafter. Every frame consists of two sub-frames. The first sub-frame is the time interval in which the policy $\pi_{\Delta}$ employs action $a=0$, and the second sub-frame is the time interval in which the policy $\pi_{\Delta}$ employs action $a=1$ until the $i$th request gets served. Note that the length of the first sub-frame is $\Delta-1$, and we denote the length of the second sub-frame by $\Delta_{1}$. Note that, $\Delta_{1}$ is geometrically distributed with the probability of success $q_{\lambda} v(i)$. 
    
    Let us consider any arbitrary frame $j$ and denote its length by $\ell$ and the total cost in that frame by $J(\Delta,\tau)$.  Then, the expected length of frame $j$ is
    \begin{align}\label{eq:32}
        \mathbb{E}[\ell] = \Delta -1 +  \frac{1}{q_{\lambda}v(i)}. 
    \end{align}
    The expected total cost in frame $j$ is
    \begin{align}\label{eq:33}
        \mathbb{E}[J(\Delta,\tau)] = \frac{\tau |\mathcal{G}(i)|}{q_{\lambda}v(i)} + \frac{1}{2}\bigg( &(\Delta-1)^{2} + \frac{2 (\Delta-1)}{q_{\lambda} v(i)}  \nonumber\\& + \frac{2-q_{\lambda}v(i)}{(q_{\lambda} v(i))^{2}}\bigg) + \frac{\mathbb{E}[\ell]}{2}.
    \end{align}
     For every request $i\in\mathcal{R}$, we have $q_{\lambda} v(i)>0$. Thus, for a fixed and finite $\Delta$, we have $\mathbb{E}[\ell]<\infty$. Now, using the renewal reward theorem, we have,
     \begin{align}
         J_{\textrm{avg}}(\Delta,\tau) = \frac{\mathbb{E}[J(\Delta,\tau)]}{\mathbb{E}[\ell]},
     \end{align}
     where we use $J_{\textrm{avg}}(\Delta,\tau)$ to denote the total average cost for policy $\pi_{\Delta}$. From (\ref{eq:32}) and (\ref{eq:33}), we have,
     \begin{align}
         J_{\textrm{avg}}(\Delta,\tau) = \frac{\Delta}{2} + \frac{1}{2 q_{\lambda}v(i)}  + \frac{\tau |\mathcal{G}(i)| +\frac{1-q_{\lambda}v(i)}{2q_{\lambda}v(i)} }{q_{\lambda}v(i)(\Delta-1)+1}.\label{eq:n35}
     \end{align}
     Note that, $J_{\textrm{avg}}(\Delta,\tau)$ is a convex function of $\Delta$, and gets minimized at $\Delta^{*}(\tau)$, where
     \begin{align}
         \Delta^{*}(\tau) = \max\bigg\{1,1+ \frac{\sqrt{1-q_{\lambda}v(i) + 2 q_{\lambda}v(i) \tau |\mathcal{G}(i)|}-1}{q_{\lambda} v(i)}\bigg\}.\label{eq:n36}
     \end{align}
     Thus, the optimal $\Delta$ for a given $\tau$ is either $\Delta^{*}_{1}(\tau) = \lfloor{\Delta^{*}(\tau)\rfloor}$ or $\Delta^{*}_{2}(\tau) = \lceil{\Delta^{*}(\tau)\rceil}$. Now, note that if $\tau\rightarrow \infty$, then $\Delta^{*}(\tau)\rightarrow \infty$. Now, from the definition of $\pi_{\Delta}$ and $\mathcal{P}_{i}(\tau)$, we say that $\mathcal{P}_{i}(\infty)= \mathcal{S}_{i}$. Note that, $\Delta_{1}^{*}(\tau)$ and $\Delta_{2}^{*}(\tau)$ are both non-decreasing functions of $\tau$. Thus, for $\tau_{2}\leq \tau_{1}$ let us assume that $\Delta_{1}^{*}(\tau_{2})<\Delta_{1}^{*}(\tau_{1})$ and $\Delta_{2}^{*}(\tau_{2}) < \Delta_{2}^{*}(\tau_{1})$. Thus, again from the definition of $\pi_{\Delta}$, it is immediate that $\mathcal{P}_{i}(\tau_{2})\subseteq\mathcal{P}_{i}(\tau_{1})$. Now, assume that $\Delta_{1}^{*}(\tau_{2})=\Delta_{1}^{*}(\tau_{1})$, $\Delta_{2}^{*}(\tau_{2}) = \Delta_{2}^{*}(\tau_{1})$, and $\Delta_{1}^{*}(\tau_{2})$ is an optimal threshold for the Lagrangian multiplier $\tau_{2}$. Then, for the Lagrangian multiplier $\tau_{1}$, an optimal threshold is always greater than or equal to $\Delta_{1}^{*}(\tau_{2})$. Thus, for this case, we have $\mathcal{P}_{i}(\tau_{2})\subseteq \mathcal{P}_{i}(\tau_{1})$. Now, let us assume that $\Delta_{1}^{*}(\tau_{2})=\Delta_{1}^{*}(\tau_{1})$, $\Delta_{2}^{*}(\tau_{2}) = \Delta_{2}^{*}(\tau_{1})$, and $\Delta_{2}^{*}(\tau_{2})$ is an optimal threshold for the Lagrangian multiplier $\tau_{2}$. In other words, we have
     \begin{align}
         & J_{\textrm{avg}}(\Delta_{2}^{*}(\tau_{2}),\tau_{2}) \leq  J_{\textrm{avg}}(\Delta_{1}^{*}(\tau_{2}),\tau_{2}) \\ &\frac{\Delta_{2}^{*}(\tau_{2})}{2} + \frac{\tau_{2}|\mathcal{G}(i)| + \frac{1-q_{\lambda}v(i)}{2q_{\lambda}v(i)}}{q_{\lambda} v(i)(\Delta_{2}^{*}(\tau_{2})-1)+1}  \nonumber\\& \leq \frac{\Delta_{1}^{*}(\tau_{2})}{2} + \frac{\tau_{2}|\mathcal{G}(i)| + \frac{1-q_{\lambda}v(i)}{2q_{\lambda}v(i)}}{q_{\lambda} v(i)(\Delta_{1}^{*}(\tau_{2})-1)+1}.\label{eq:38}
     \end{align}
     Let us define $\epsilon = \tau_{1}-\tau_{2}$. Note that, $\Delta_{2}^{*}(\tau_{2})\geq \Delta_{1}^{*}(\tau_{2})$. Thus, from (\ref{eq:38}), we have
     \begin{align}
         &\frac{\Delta_{2}^{*}(\tau_{2})}{2} + \frac{(\tau_{2}+\epsilon)|\mathcal{G}(i)| + \frac{1-q_{\lambda}v(i)}{2q_{\lambda}v(i)}}{q_{\lambda} v(i)(\Delta_{2}^{*}(\tau_{2})-1)+1} \nonumber\\& \leq \frac{\Delta_{1}^{*}(\tau_{2})}{2} + \frac{(\tau_{2}+\epsilon)|\mathcal{G}(i)| + \frac{1-q_{\lambda}v(i)}{2q_{\lambda}v(i)}}{q_{\lambda} v(i)(\Delta_{1}^{*}(\tau_{2})-1)+1}.\label{eq:40}
     \end{align}
     Now, as we assumed that $\Delta_{1}^{*}(\tau_{2}) = \Delta_{1}^{*}(\tau_{1})$ and $\Delta_{2}^{*}(\tau_{2}) = \Delta_{2}^{*}(\tau_{1})$, from (\ref{eq:n35}) and (\ref{eq:40}), we have
     \begin{align}
         J_{\textrm{avg}}(\Delta_{2}^{*}(\tau_{1}),\tau_{1}) \leq J_{\textrm{avg}}(\Delta_{1}^{*}(\tau_{1}),\tau_{1}). 
     \end{align}
     Thus, for the Lagrangian multiplier $\tau_{1}$, $\Delta_{2}^{*}(\tau_{1})=\Delta_{2}^{*}(\tau_{2})$ is an optimal threshold. Thus, for this case also, we have $\mathcal{P}_{i}(\tau_{2}) \subseteq \mathcal{P}_{i}(\tau_{1})$. Now, consider the cases where $\Delta_{1}^{*}(\tau_{2})=\Delta_{2}^{*}(\tau_{2})=\Delta_{1}^{*}(\tau_{1})<\Delta_{2}^{*}(\tau_{1})$ and $\Delta_{1}^{*}(\tau_{2})<\Delta_{1}^{*}(\tau_{1})=\Delta_{2}^{*}(\tau_{2})=\Delta_{2}^{*}(\tau_{1})$. In both the cases, it is immediate that $\mathcal{P}_{i}(\tau_{2})\subseteq\mathcal{P}_{i}(\tau_{1})$. We have explored all possible cases, and for all of them we have proved $\mathcal{P}_{i}(\tau_{2})\subseteq\mathcal{P}_{i}(\tau_{1})$, which completes this proof.
\end{ProofOfTh1}
    
\section{Proof of Theorem~\ref{th:4}}\label{appen:th2}
For a state $s\in\mathcal{S}_{i}$, first we find $\tau_{\textrm{eq}}(s)$, which we define as
\begin{align}\label{eq:41}
    \tau_{\textrm{eq}}(s) = \inf\{\tau \geq 0| J_{\textrm{avg}}(s,\tau) =J_{\textrm{avg}}(s+1,\tau) \}.
\end{align}
If there does not exist any $\tau \geq 0$, such that $J_{\textrm{avg}}(s,\tau) =J_{\textrm{avg}}(s+1,\tau)$, then we define $\tau_{\textrm{eq}}(s)=\infty$. Now, solving for $J_{\textrm{avg}}(s,\tau) = J_{\textrm{avg}}(s+1,\tau)$, we get
\begin{align}
    \tau_{\textrm{eq}}(s) = \frac{s(q_{\lambda}v(i) s - q_{\lambda}v(i) +2)}{2 \lambda}.
\end{align}
Note that $\tau_{\textrm{eq}}(s)$ is unique, positive and finite. Now, for a state $s$, we define $\tau^{*}(s)$ to be the Lagrangian multiplier that satisfies,
\begin{align}
    s= \Delta^{*}(\tau^{*}(s)). 
\end{align}
Thus, for a state $s\in\mathcal{S}_{i}$, we have 
\begin{align}
    \tau^{*}(s) = \frac{q_{\lambda}v(i) (s-1)^{2} + 2(s-1) +1}{2 |\mathcal{G}(i)|}.
\end{align}
It is immediate that 
\begin{align}\label{eq:45}
    \tau_{\textrm{eq}}(s-1)<\tau^{*}(s) < \tau_{\textrm{eq}}(s).
\end{align}
For $\epsilon_{1}>0$, from (\ref{eq:n36}) and from the definition of $\tau^{*}(s)$, we have,
\begin{align}
    \Delta^{*}(\tau^{*}(s)-\epsilon_{1}) \leq s.
\end{align}
Thus, for a Lagrangian multiplier $\tau^{*}(s)-\epsilon_{1}$, the optimal threshold is less than or equal to $s$, which translates to $s\notin\mathcal{P}_{i}(\tau^{*}(s)-\epsilon_{1})$. From (\ref{eq:45}) and (\ref{eq:n36}), we say that for $\tau^{*}(s)\leq\tau\leq \tau^{*}(s+1)$ the optimal threshold lies between $s$ and $s+1$. Now, for $\epsilon_{1}>0$, from (\ref{eq:n35}) and (\ref{eq:41})  we have,
\begin{align}
    J_{\textrm{avg}}(s,\tau^{\textrm{eq}}(s)-\epsilon_{1}) \leq  J_{\textrm{avg}}(s+1,\tau^{\textrm{eq}}(s)-\epsilon_{1}).
\end{align}
Thus, for $\tau^{*}(s)\leq\tau< \tau_{\textrm{eq}}(s)$, from (\ref{eq:41}) we have $s\notin \mathcal{P}_{i}(\tau)$. For $\tau=\tau_{\textrm{eq}}(s)$, from (\ref{eq:41}) we have $s\in\mathcal{P}_{i}(\tau)$, which concludes this proof.

\bibliographystyle{unsrt}
\bibliography{references}

\end{document}